\documentclass[a4paper]{aa}  
\usepackage{natbib}
\usepackage{graphicx}
\usepackage{epsfig}
\usepackage{amsmath}
\usepackage{txfonts}
\begin{document}
   \title{Integrated specific star formation rates of galaxies,
   groups, and clusters: A continuous upper limit with stellar
   mass?\thanks{Based on observations collected at the Centro
   Astron\'omico Hispano Alem\'an (CAHA), operated by the
   Max-Planck-Institut f\"ur Astronomie, Heidelberg, jointly with the
   Spanish National Commission for Astronomy.}$^,$\thanks{Based on
   observations collected at the European Southern Observatory, Chile;
   ESO proposals 63.O-0005, 64.O-0149, 64.O-0158, 64.O-0229,
   64.P-0150, 65.O-0048, 65.O-0049, 66.A-0123, 66.A-0129, 66.A-0547,
   68.A-0013, 69.A-0014, and LP168.A-0485.}}

   \authorrunning{G.~Feulner, U.~Hopp, \& C.~S.~Botzler}

   \titlerunning{Integrated specific star formation rates of galaxies,
   groups, and clusters}

   \author{Georg Feulner\inst{1,2},
	   Ulrich~Hopp\inst{1,2},
	   \and
	   Christine~S.~Botzler\inst{1,3,4}}

   \offprints{G. Feulner, \email{feulner@usm.lmu.de}}

   \institute{Universit\"ats--Sternwarte M\"unchen, Scheinerstra\ss e
              1, D--81679 M\"unchen, Germany
         \and
	 Max--Planck--Institut f\"ur
              extraterrestrische Physik, Giessenbachstra\ss e 1,
              D--85748 Garching, Germany
	 \and
	 University of Auckland, Private Bag 92019, Morrin Road, 
	 Glen Innes, Auckland, New Zealand
	 \and
	 University of Canterbury, Private Bag 4800, Christchurch, 
	 New Zealand
}
   \date{Received 24 February 2006; accepted 22 March 2006}

 
  \abstract
{} 
%
{We investigate the build-up of stellar mass through star formation in
  field galaxies, galaxy groups, and clusters in order to better
  understand the physical processes regulating star formation in
  different haloes.}
%
{In order to do so we relate ongoing star formation activity to the
  stellar mass by studying the integrated specific star formation rate
  (SSFR), defined as the star-formation rate per unit stellar mass, as
  a function of integrated stellar mass for samples of field galaxies,
  groups of galaxies, and galaxy clusters at $0.18 \le z \le 0.85$. The
  star formation rate (SFR) is derived from the ultraviolet continuum
  for the galaxies and group members, and from emission line fluxes
  for the cluster galaxies. The stellar masses are computed from
  multi-band photometry including the near-infrared bands for the
  galaxies and groups, and from the dynamical mass for the cluster
  sample.}
%
{For the first time, integrated SSFRs for clusters and groups are
  presented and related to the SSFRs of field
  galaxies. Tentatively, we find a continuous upper limit for
  galaxies, groups, and clusters in the SSFR-stellar mass plane over
  seven orders of magnitude in stellar mass. This might
  indicate that the physical processes which control star formation
  in dark matter haloes of different mass have the same scaling with
  mass over a wide range of masses from dwarf galaxies to massive
  clusters of galaxies.}
%
%
{}


   \maketitle
%

\section{Introduction}
\label{sec:intro}

In 1996, \citeauthor{Cowie1996} investigated the contribution of star
formation to the build-up of stellar mass for different galaxy
masses. They found that at higher redshifts a population of massive,
heavily star forming galaxies emerges which cannot be found in the
local universe, a phenomenon they termed ``down-sizing''. This, in
turn, implies that the more massive galaxies found in today's universe
have older stellar populations. After this pioneering work the
specific star formation rate (SSFR), defined as the star formation
rate (SFR) per unit stellar mass, was used to study this connection
and to follow its evolution with redshift. It is now well established
that galaxies show an SSFR falling with stellar mass with a clear
upper limit in SSFR \citep{Guzman1997, BE00, PerezGonzalez2003,
Brinchmann2004, Fontana2003, Bauer2005, Bell2005}, with a fraction of
strongly star-forming galaxies at high SSFRs \citep{Bell2005,
Hammer2005, Perez2005}. Moreover, the most massive galaxies are
dominated by the oldest stellar populations \citep{fdfmf, munics7} and
show a marked increase of their mean SSFR around $z \sim 2$
\citep{fdfssfr, Juneau2005}.

Similar studies on the star formation rate in clusters of galaxies
have been carried out during the last decade finding a decrease of the
integrated SFR per unit dynamical mass with mass \citep{Finn2004,
Finn2005} or, in some sense equivalently, a falling fraction of
[OII]-emitting galaxies with velocity dispersion
\citep{Poggianti2006}. For the cluster members, evidence was found for
a down-sizing scenario similar to the field \citep[e.g.][]{Smail1998,
Tanaka2005, Poggianti2006}.

For galaxy groups, systematic studies are still rare, although several
studies of star-formation activity as a function of environment have
been performed \citep{Hashimoto1998, Balogh2004,
Wilman2005}. Recently, \citet{Weinmann2006} presented SSFRs for
galaxies in SDSS groups showing a decline of the SSFR with halo mass.

In this work we try to combine information on the SSFR of field
galaxies with the integrated SSFR of galaxy groups and clusters to
investigate the build-up of stellar mass in haloes over a wide range
of masses from dwarf galaxies to massive clusters. Studies
like these are important to better constrain the physical processes
responsible for controlling star formation in different environments
and thus haloes of different masses.

This Letter is organised as follows. In Sect.~\ref{sec:samples} we
describe the galaxy, group, and cluster samples as well as our methods
to derive SFRs and stellar masses. Section~\ref{sec:ssfr} presents our
results on the distribution of the different samples in the SSFR-stellar
mass plane, before we discuss and summarise our work in
Sect.~\ref{sec:summ}.  Throughout this Letter we assume a concordance
cosmology with $\Omega_M = 0.3$, $\Omega_\Lambda = 0.7$ and $H_0 = 70
\, \mathrm{km} \, \mathrm{s}^{-1} \, \mathrm{Mpc}^{-1}$. All
magnitudes are given in the Vega system.

\section{Deriving star formation rates and stellar masses for the different 
samples}
\label{sec:samples}

\subsection{The field galaxy sample}

The field galaxies used in this study are taken from a combined sample
derived from the FORS Deep Field \citep{fdf1, fdflf1} and the GOODS-S
field. It is the same sample already used and discussed in
\citet{fdfssfr}. The FDF offers photometry in the $U$, $B$, $g$, $R$,
$I$, 834~nm, $z$, $J$ and $K$ bands and is complimented by deep
spectroscopic observations \citep{fdfspec}. In this Letter we use the
$I$-selected sub-sample covering the deep central part of the field
($\sim$~40 arcmin$^2$) as described in \citet{fdflf1}, containing 5557
galaxies down to $I=26.4$ (50\% completeness limit for point
sources). Photometric redshifts of FDF galaxies have an
accuracy of $\Delta z / (1+z) = 0.03$ \citep{fdflf1}.\\
Our $K$-band selected catalogue for the GOODS-S field
\citep{Salvato2006} is based on the publicly available 8
$2.5\times2.5$ arcmin$^2$ $J$, $H$, and $K_s$ VLT/ISAAC images
complimented by observations in $U$, $B$, $V$, $R$, and $I$. The
sample contains 3237 galaxies with a photometric redshift accuracy of
$\Delta z / (1+z) = 0.05$ down to $K=23$ over a field of view of
$\sim$~50 arcmin$^2$.\\
For both fields, SFRs are derived from the luminosity $L_{1500}$ of
the ultraviolet continuum at $\lambda \simeq 1500$\AA\ \citep{fdfsfr}
converting it to an SFR $\dot{\varrho}_{\,\ast}$ (in units of solar
masses per year) as described in \citet{Madau1998} and assuming a
Salpeter initial mass function \citep{Salpe55}

\begin{equation}
\dot{\varrho}_{\,\ast} \; = \; 1.25 \times 10^{-28} \: {\cal
M}_{\,\odot} \: \mathrm{yr}^{-1} \: \frac{L_{1500}}{\mathrm{erg} \,
\mathrm{s}^{-1} \, \mathrm{Hz}^{-1}} ,
\end{equation}

while stellar masses ${\cal M}_{\,\ast}$ (in units of solar masses)
are computed from fitting stellar population synthesis models to the
galaxies' broad-band photometry \citep{masscal}. A standard
correction for dust extinction is applied to the SFR following the
recipe of \citet{Hopkins2004}. The SSFR ${\cal S}$ is then simply
calculated from

\begin{equation}
 {\cal S} \; \equiv \; \frac{\dot{\varrho}_{\,\ast}}{{\cal
 M}_{\,\ast}} .
\end{equation}

To ensure fair comparison with the cluster sample described below we
have restricted the redshift range for the field galaxies to $0.18 \le
z \le 0.85$ yielding 2898 objects with an average redshift of
$\langle z \rangle = 0.54$. Note that within this $z$ interval the
fraction of undetected dusty star-forming galaxies is still small
\citep{Franceschini2003}.

\subsection{The galaxy group sample}

The integrated SSFRs and stellar masses for the groups are based on
the Munich Near-Infrared Cluster Survey (MUNICS; \citealt{munics1,
munics5}), a wide-area, medium deep photometric and spectroscopic
survey in the $B$, $V$, $R$, $I$, $J$, and $K$ bands covering an area
of about 0.3 square degrees down to $K \simeq 19$ and $R \simeq 24$
(50\% completeness limit for point sources). Group membership on the
photometric redshift catalogue with its accuracy of $\Delta z / (1+z)
= 0.06$ is assigned according to a modified version of the
friends-of-friends algorithm, specifically designed to cope with
photometric redshift datasets \citep{Botzler2004}. The resulting
structure catalogue is presented in \citet{munics8} and comprises 162
structures (mostly groups) containing 890 galaxies in total.\\
SFRs and stellar masses for the individual group members are computed
in the same manner as for the field galaxies described above (see also
\citealt{munics7}) and summed for each group, resulting in what we
call \textit{integrated} values. To ensure fair comparison with the
cluster sample described below we have restricted the redshift range
for the groups to $0.18 \le z \le 0.85$ leaving us with 137 groups
containing 710 galaxies with an average redshift of $\langle z
\rangle = 0.50$.

\subsection{The cluster sample}

Since the survey volume probed by MUNICS is too small to
contain massive clusters, integrated SFRs and stellar masses for
galaxy clusters are obtained from the sample described in
\citet{Finn2004, Finn2005} who derive SFRs from the H$\alpha$ line
emission. The sample contains 8 clusters at $0.18 \le z \le 0.85$ with
SFR measurements of galaxies, usually within the virial
radius. The average redshift of the sample is $\langle z
\rangle = 0.53$. Their integrated SFR values are corrected for dust
extinction using $A_{H\alpha} = 1$. Since no measurements for
the total stellar mass of the clusters are available, we compute this
quantity from the dynamical mass (derived from the cluster velocity
dispersion). To accomplish this we make use of the relation between
the stellar mass ${\cal M}_{\,\ast}$ and the dynamical mass ${\cal
M}_{\,\mathrm{500}}$ within the radius ${\cal R}_{\,\mathrm{500}}$
(within which the mean density is 500 times the critical density)
derived by \citet{Lin2003} from $K$-band observations of local
clusters:

\begin{equation}
\frac{{\cal M}_\ast}{{\cal M}_{\,\mathrm{500}}} \; = \; (1.64 \pm 0.10)
\times 10^{-2} \; \left( \frac{{\cal M}_{\,\mathrm{500}}}{3 \times 10^{14}
{\cal M}_{\,\odot}} \right)^{-0.26 \pm 0.09}
\end{equation}

To convert the values of ${\cal M}_{\,\mathrm{200}}$ given in
\citet{Finn2005} to ${\cal M}_{\,\mathrm{500}}$ we use the following
relation for the dynamical mass ${\cal M}_{\,\delta}$ as a
function of the density contrast $\delta$ \citep{Horner1999}:

\begin{equation}
{\cal M}_{\,\delta} \: \propto \: \delta\,^{-0.266 \: \pm \: 0.022} \;\;\; 
\mathrm{or} \;\;\;
{\cal M}_{\,\mathrm{500}} \: \simeq \: 0.78 \: {\cal M}_{\,\mathrm{200}}
\end{equation}

Estimating the integrated SFR and stellar mass of galaxy
clusters in a different manner than for the galaxies and groups is, of
course, not optimal; for future studies it would be desirable to
derive the SFRs and the masses using the same methods.

\section{Star formation and stellar mass in field galaxies, groups, and 
clusters}
\label{sec:ssfr}

\begin{figure}
\begin{center}
\epsfig{figure=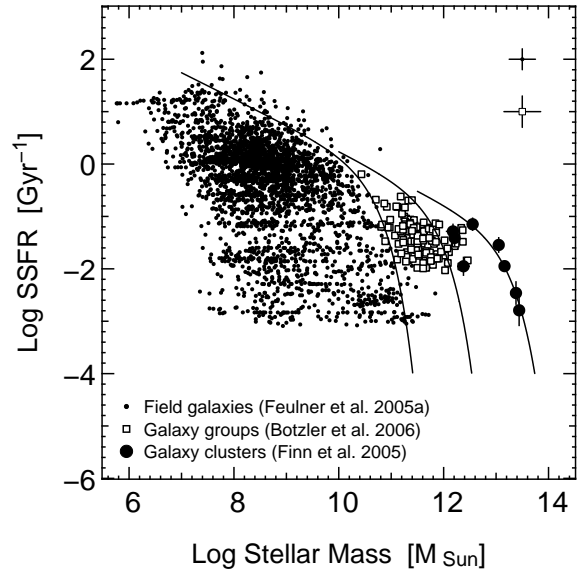,width=0.85\columnwidth}
\caption{Integrated SSFR versus stellar mass for field galaxies (small
  filled circles), galaxy groups (open squares), and galaxy clusters
  (large filled circles) in the redshift range $0.18 \le z \le
  0.85$. The error crosses in the upper left corner give conservative
  estimates of the errors for the field galaxies and the groups, whereas
  individual error bars are attached to the cluster values. The solid
  lines show approximate limits to the point distributions (see text
  for details).}
\label{fig:ssfr}
\end{center}
\end{figure}

In Fig.~\ref{fig:ssfr} we present the resulting integrated SSFR versus
stellar mass diagram for field galaxies, groups, and clusters of
galaxies. For groups and clusters, this quantity has -- to our
knowledge -- not been presented before, while it has been previously
shown for field galaxies \citep[e.g.][ and references
therein]{fdfssfr}. For the first time the distribution of field
galaxies, groups, and clusters in this diagnostic diagram can be
studied in context.

It is now well established that the SSFR of field galaxies is
decreasing with increasing stellar mass. This trend holds up to
stellar masses of $\log {\cal M}_{\,\ast} / {\cal M}_{\,\odot} \simeq
11$ where we reach the high-mass cut-off of the stellar mass function
in the redshift range $0.18 \le z \le 0.85$
\citep[e.g.][]{Fontana2004, fdfmf}.

Interestingly, the integrated SSFRs of groups and clusters continue
this trend to higher stellar masses.  The limiting SSFR of the groups
from MUNICS follow the same sequence up to $\log {\cal M}_{\,\ast} /
{\cal M}_{\,\odot} \simeq 12.5$, reaching the SSFRs of clusters with
$\log {\cal M}_{\,\ast} / {\cal M}_{\,\odot} \lesssim 13.5$ at the
very end of this sequence. Groups and clusters seem to form a
natural extension of the SSFR distribution of galaxies, with the upper
limit shaping a continuous sequence over at least seven orders of
magnitude in stellar mass. It is not surprising that there is some
overlap between very massive galaxies and poor groups as well as
between massive groups and poor clusters. The slightly smaller values
for the integrated SSFRs of the MUNICS groups could be attributed to
the selection of the group members in photometric redshift space; the
algorithm might miss some members in the outer regions more likely to
be star-forming galaxies. Note also that due to the near-infrared
selection the MUNICS group sample may be biased against high-SFR
galaxies with low dust attenuation.

For easier analysis, the approximate upper boundaries to the SSFR
${\cal S}$ for the different samples in Fig.~\ref{fig:ssfr} can be
described by the following functional form (similar to the Schechter
parametrisation of the luminosity function, \citealt{Schechter76}):

\begin{equation}
 {\cal S} \; = \; {\cal S}_{\,0} \: 10^{\,{\cal
 M}_{\,\ast}\:(1\,+\,\alpha)} \: \exp \left( \, 10^{\,{\cal
 M}_{\,\ast}\,-\, {\cal M}_{\,0}} \, \right)
\end{equation}

The free parameters of this function describe the normalisation
(${\cal S}_{\,0}$), the location of the break (${\cal M}_{\,0}$), and
the slope at lower stellar masses ($\alpha$). Their approximate values
are $(\log {\cal S}_{\,0} / \mathrm{Gyr}^{-1}, \: \log {\cal M}_{\,0}
/ {\cal M}_{\,\odot}, \: \alpha) \simeq (5.25, \: 10.5, \: -1.5)$ for
the field galaxies, $(5.25, \: 11.7, \: -1.5)$ for the groups, and
$(5.25, \: 13.0, \: -1.5)$ for the clusters. The slope $\alpha$ can be
derived from the field-galaxy sample only but seems to apply also to
the other samples. The curves corresponding to these values are
plotted in Fig.~\ref{fig:ssfr}. Note, again, that there is a smooth
transition from groups to clusters, so the upper mass limit for groups
should be taken with a grain of salt.

\nocite{fdfssfr, munics8, Finn2005}

\section{Discussion and conclusions}
\label{sec:summ}

In this Letter we have for the first time presented the integrated
SSFR for groups and clusters of galaxies and compared it to the SSFR
of the field galaxy population. Moreover, we tentatively find
a continuous upper limit for galaxies, groups, and clusters in the
SSFR-stellar mass plane over seven orders of magnitude in stellar
mass. This might indicate that the processes which control
star formation in dark matter haloes of different mass have the same
scaling with mass over a wide range of masses from dwarf galaxies to
massive clusters of galaxies.

The physical processes responsible for the ``down-sizing'' phenomenon
witnessed in \textit{individual} galaxies are not yet well
understood. An early formation epoch for massive galaxies or ``dry
merging'' \citep{Faber2006, Bell2006} of lower-mass galaxies as well
as quenching of star formation in more massive haloes by feedback
mechanisms \citep[e.g.][]{Scannapieco2005} are among the discussed
possibilities. Of course, we could also see the result of a
combination of these processes or be faced with different evolutionary
paths leading to the population of massive galaxies with old stellar
populations.

The fact that the \textit{integrated} SSFRs of groups and clusters of
galaxies seem to continue the trend displayed by the field galaxy
population towards higher masses is intriguing. For the
\textit{individual} galaxies within these structures, one can
naturally expect a similar general behaviour as for their counterparts
in the field, modified by environmental effects. It has been known for
a long time that higher density environments are occupied by galaxies
with morphologically earlier types \citep[e.g.][]{Dressler1980,
Postman1984, Dressler1997} and with overall redder colours (and thus
lower star-formation activity; \citealt{BO78a}). Moreover, ellipticals
in higher-density environments are on average older than their
low-density counterparts \citep{Thomas2005}, and star formation
activity in groups seems to be lower than in the field
\citep{Wilman2005}. But the fact that the upper limit of the
\textit{integrated} SSFRs of all these objects, from dwarf galaxies to
rich clusters, seems to follow a continuous sequence in the
SSFR--stellar mass plane seems to suggest that there could be
a smooth transition from the field to the clusters, which in turn
might imply that the physical processes responsible for the
lower integrated star formation activity in higher mass
haloes are the same over this wide range of stellar masses, or at
least have the same scaling with stellar mass.

The analysis presented here is made possible by the
availability of large samples of field galaxies with well studied
properties, and by the advent of group and cluster catalogues with
photometric and spectroscopic data for large number of
members. However, statistics is still rather poor for groups and
clusters, and  we could not derive SFRs and stellar
masses using the same methods in all samples. Future studies of large
and homogeneous samples of groups, clusters and their member galaxies
will result in progress in the study of galaxy evolution as
a function of local density, and allow us to better constrain the
physical processes responsible for controlling star formation in
different environments and thus haloes of different masses.

\begin{acknowledgements}
The authors thank R.~Bender and C.~Mendes de Oliveira for helpful
suggestions, J.~Snigula for assistance with the structure catalogue,
N.~Drory for making his mass-fitting code available, A.~Gabasch as
well as M.~Salvato for their work on FDF and GOODS-S, and the
anonymous referee for his comments. G.F.\ and C.S.B.\ acknowledge
funding by the DFG, G.F.\ also by the MPG.
\end{acknowledgements}

\bibliographystyle{aa}
\bibliography{mnrasmnemonic,literature}

\end{document}